\documentstyle[12pt,epsfig,axodraw]{article}
\newcommand{\MeV}{{\rm MeV}}
\newcommand{\GeV}{{\rm GeV}}

\newcommand{\fm}{{\rm fm}}
\renewcommand{\Im}{{\rm Im}}
\renewcommand{\Re}{{\rm Re}}

\newcommand{\lsim}{$\raisebox{-0.8ex} {$\stackrel{\textstyle <}{\sim}$}$}
\newcommand{\gsim}{$\raisebox{-0.8ex} {$\stackrel{\textstyle >}{\sim}$}$}

\begin{document}
\title{Modification of the $\phi$-meson spectrum \newline in nuclear matter
\thanks{Work supported in part by GSI and BMBF}}
\author{F. Klingl, T.Waas and W. Weise\\ Physik-Department\\ Technische Universit\"at
M\"unchen\\ D-85747 Garching, Germany}
\date{}
\maketitle
\begin{abstract}
  The vacuum spectrum of the $\phi$-meson is characterized by its decay into
  $K{\overline K} $. Modifications of the $K {\overline K}$-loops in baryonic
  matter change this spectrum. We calculate these in-medium modifications
  taking both s- and p-wave kaon-nucleon interactions into account. We use
  results of the in-medium $K$ and $\overline K$ spectra determined
  previously from a coupled channel approach based on a chiral effective
  Lagrangian. Altogether we find a very small shift of the $\phi$
  meson mass, by less than 10 MeV at normal nuclear matter density $\rho_0$. The
  in-medium decay width of the $\phi$ meson increases such that its life time
  at $\rho=\rho_0$ is reduced to less than 5 fm/c. It should
  therefore be possible to observe medium effects in reactions such as $\pi^- p
  \to \phi n$ in heavy nuclei, where the $\phi$ meson can be produced with
  small momentum.
\end{abstract}

{\bf Introduction.} The study of in-medium properties of hadrons is a topic of
continuing interest.  Experiments planned at GSI (HADES) and running at CERN
(e.g. CERES) detect dilepton pairs in high energy collisions of nuclei. This
opens the possibility of investigating vector mesons in hot
and dense hadronic matter. In order to explore such medium effects it is
necessary that the vector mesons decay inside the hot and dense region of the
collision zone. The $\phi$ meson with its small width of 4.4 MeV has a lifetime of
about 45 fm/c which is obviously too large to observe any medium effects. On
the other hand, we demonstrate in this note that medium modifications are
expected to increase the $\phi$
width and shorten its lifetime to less than 5 fm/c at the density of normal nuclear
matter. One then enters the range in which medium effects of slowly moving
$\phi$-mesons could become visible. A reaction of particular interest is $\pi^-
p \to \phi n$ in heavy nuclei. This process violates the OZI rule but
substantial $\omega \phi$ mixing makes the $\phi$ production rate large enough
to be well detectable. Such an experiment could be performed at GSI where a
pion beam can be used in combination with HADES to measure $\pi ^- p \to
\omega n$ \cite{1}. 

The purpose of this paper is to provide a systematic calculation of the
in-medium $\phi$ meson self energy. We first review briefly the vacuum
properties of the $\phi$ meson and discuss its self-energy. We then extend this
for finite densities by including the in-medium interactions of the decay
kaons, taking into account both s- and p-wave interactions of $K$ and
${\overline K}$ with nucleons in nuclear matter. In the final part a brief summary and
discussion of the results will be given.

{\bf \label{2} The $\phi$ meson in vacuum.} The $\phi$ meson is observed as a
pronounced resonance in the strange quark sector of the electromagnetic
current-current correlation function \cite{9}. The Fourier transform of the correlation
function is 
\begin{equation}
  \Pi_{\mu\nu}(q)=i\int d^4x \: e^{iq\cdot x}\langle
  0|{\cal T}j_{\mu}(x) j_{\nu}(0)|0\rangle  ,
 \label{2.1}
\end{equation} 
where in the present context $j_\mu$ represents the strange quark current,
\begin{equation}
  \label{2.2}
  j_{\mu}=-\frac{1}{3}(\bar{s}\gamma_{\mu}s). 
\end{equation}
Current conservation leads to a transverse tensor structure
\begin{equation}
  \Pi_{\mu\nu}(q)=\left(g_{\mu\nu}-\frac{q_{\mu}q_{\nu}}{q^2}\right)\Pi (q^2),
\label{2.3}
\end{equation}
which defines the scalar function $ \Pi (q^2)=-\frac{1}{3} \Pi_\mu^\mu(q)$. The low energy spectrum
of the correlation function  is well described by Vector Meson
Dominance (VMD). We use our improved VMD approach of ref.\cite{3} which gives
\begin{equation}
  \label{2.4}
  \Im \Pi (q^2)= \frac{\Im \Pi_\phi^{\rm vac}(q^2)}{g_\phi^2} \left| \frac{(1-a_\phi)\, q^2-\stackrel{ \rm o }{m}_\phi^2}{q^2-\stackrel{ \rm o
   }{m}_\phi^2-\Pi_\phi^{\rm vac}(q^2)}\, \right|^2.
\end{equation}
Here we have introduced the bare mass $\stackrel{ \rm o }{m}_\phi$ of the $\phi$ meson,
$g_\phi=-3g/\sqrt{2} \simeq -14$ is its strong coupling constant, and $a_\phi$ a
constant which describes deviations from universality of the $\phi\rightarrow
e^+e^-$ and $\phi\rightarrow K\overline K$ couplings. This constant is close to
unity, i.e. deviations from universality are small. For our present purpose extreme
fine-tuning is not necessary 
and we can set $a_\phi=1$. 

The self-energy $\Pi_\phi^{\rm vac}$ of the
$\phi$ meson in the vacuum consists of three parts:
\begin{equation}
 \label{2.5}
  \Pi_\phi^{\rm vac} = \Pi_{\phi \to K^+ K^-}^{\rm vac} + \Pi_{\phi \to K^0_L K^0_S}^{\rm vac}+ \Pi_{\phi \to
  3 \pi}^{\rm vac}\; ,
\end{equation}
describing the coupling of the $\phi$ to the $K {\overline K}$ and three-pion
channels.  The last term violates the OZI rule, but despite the small $\omega
\phi$ mixing angle it contributes about 15 percent to the total $\phi$ meson
decay width. We include its imaginary part as given in ref.\cite{3} but 
focus here on the more important parts of the self-energy coming from the decay
into $K{\overline K}$ channels. They are related by SU(3) to the $\rho -\pi \pi $
self energy and can be written in the form of a one-loop integral \cite{3,4}:
\begin{equation}
\label{2.6}
\Pi_{\phi \to K^+K^-}^{\rm vac}(q^2) =\frac{-i g^2}{6} \int \frac{d^4l}{(2\pi)^4} \left[ \frac{(2l-q)^2}{(l^2-m_{K}^2+i\epsilon)((l-q)^2-m_{K}^2+i\epsilon)}-\frac{8}{l^2-m_{K}^2+i\epsilon}\right].
\end{equation} 
The first term of the integrand involves a propagating $K^+K^-$ pair; the
second (tadpole) term ensures gauge invariance at the level of the hadronic
effective theory.
Here we have introduced the strong meson coupling $g=6.5$ and the 
charged kaon mass
$m_{K}=493$ MeV. Evaluating this integral and applying regularization using a
subtracted dispersion relation \cite{3} we get
\begin{eqnarray}
  \label{2.7}
  \Re\; \Pi_{\phi \to K^+K^-}^{\rm vac} (q^2)&  = &c_0 \, q^2 -\frac{g^2}{48 \pi^2} \left[q^2 {\cal G}
 (q^2,m_{K}^2)- 4 m_{K}^2\right]\; , \\
\label{2.8}
\Im \;\Pi_{\phi \to K^+K^-}^{\rm vac} (q^2) &  = &- \frac{g^2}{96\pi} q^2 \left( 1 -
\frac{4m_{K}^2}{q^2} \right) ^\frac{3}{2} \; \Theta (q^2-4m_{K}^2)\;, 
\end{eqnarray}
where the subtraction constant $c_0=0.11$ has been fixed to give a best fit to
data as explained in ref.\ \cite{3}. This
leads to a bare mass $\stackrel{ \rm o }{m}_\phi=910$
MeV in eq.(\ref{2.4}). In eq.(\ref{2.7}) we insert \cite{3}
\begin{equation}
  \label{2.9}
  {\cal G}(q^2,m^2) = \left\{ \begin{array}{r@{\quad:\quad}l} 
\left(\frac{4m^2}{q^2}-1 \right)^{\frac{3}{2}} \arcsin{\frac{\sqrt{q^2}}{2\: m}} & 0<q^2<4m^2 \\ -\frac{1}{2} 
\left(1-\frac{4m^2}{q^2} \right) ^{\frac{3}{2}} \ln{\left[
\frac{1+\sqrt{1-\frac{4m^2 }{q^2}}}{1-\sqrt{1-\frac{4m^2 }{q^2}}} \right] } &
4m^2<q^2 \; {\rm or}\; q^2<0 \end{array}
\right. . 
\end{equation}
For $ \Pi_{\phi \to K_S K_L}^{\rm vac}$ the same expressions as in
eqs.(\ref{2.7},\ref{2.8}) hold with the charged kaon mass $m_{K}$ replaced by
$m_{K^0}$. Using these self-energies as input in eq.(\ref{2.4}) we plot the
spectrum of the vacuum correlation function $\Pi(q^2)$ in Fig.4a (dashed
line). We also show the real part of the $\phi$ meson propagator $D_\phi(q^2) =
[ q^2-\stackrel{ \rm o }{m}_\phi^2-\Pi_\phi^{\rm vac}(q^2)]^{-1}$ in fig. 4b
(dashed line). The zero of $\Re\; D_\phi$ determines the physical mass of the free
$\phi$ meson.

{\bf \label{3} The $\phi$ meson in medium}. We choose a Lorentz frame with
nuclear matter at rest. In the following we consider the case with the $\phi$
meson at rest ($q=(\omega,\vec{q}=0$)), so as to determine the in-medium mass
of the $\phi$-meson. The tensor structure of the correlation function then
reduces to a term proportional to the spacelike Kronecker symbol $\delta_{ij}$.
All time components vanish, and one can single out a scalar function by taking
the trace $ \Pi =\frac{1}{3} \Pi_i^i$. The spectral function has a form
analogous to that in the vacuum \cite{5,6}. One only needs to replace $q^2$ by $\omega^2$ and
the vacuum self-energy by the in-medium self energy $\Pi_\phi(\omega^2,\,
\rho)$ of the $\phi$ meson, with
\begin{equation}
  \label{3.1}
  \Im \Pi (\omega^2,\rho)= \frac{\Im \Pi_\phi(\omega^2,\rho)}{g_\phi^2} \left| \frac{(1-a_\phi)\, \omega^2-\stackrel{ \rm o }{m}_\phi^2}{\omega^2-\stackrel{ \rm o
   }{m}_\phi^2-\Pi_\phi(\omega^2,\rho)}\, \right|^2,
\end{equation}
where we use $a_\phi=1$ again as a good approximation. The difference
between the vacuum and the in-medium self-energy defines the density
dependent effective $\phi$-nucleon amplitude $T_{\phi N}$, as follows:
\begin{equation}
  \label{3.2a}
  \rho T_{\phi N}(\omega,\rho) =  \Pi^{\rm vac}_\phi (\omega^2)- \Pi_\phi (\omega^2,\rho) .
\end{equation}
In ref.\cite{6} we have shown that to leading order in density $\rho$ this
quantity reduces to the free forward $\phi$-nucleon scattering amplitude $T_{\phi N} (
\omega)$ (at $\vec{q}=0$) and we write in this approximation:
\begin{equation}
\label{3.2}
  \Pi_\phi (\omega ,\vec{q}=0;\rho) = \Pi_\phi^{\rm vac}(\omega^2)-\rho \,
  T_{\phi N} (  \omega)+... \, ,
\end{equation}
where the dots represent terms of higher order in density.

The primary modification of the self-energy $\Pi_\phi$ comes from the
interactions of the intermediate $K$ and ${\overline K}$ mesons with nucleons in the
nuclear medium. The kaon propagators in eq.(\ref{2.6}) are then to be replaced
by the in-medium propagators,
\begin{equation}
  \label{3.5}
  \frac{1}{l^2-m_{K^\pm}^2+i\epsilon} \rightarrow
  D_{K^\pm}(l_0,\,\vec{l};\,\rho)=
  \frac{1}{l_0^2-\vec{l}^{\:2}-m_{K^\pm}^2-\Sigma_{K^\pm}(l_0,\, \vec{l};\,\rho)},
\end{equation}
where $\Sigma_{K^\pm}$ are the $K^+$ and $K^-$ self-energies in nuclear matter
(or, correspondingly, those of $K_0$ and ${\overline K}_0$ where applicable). The
kaon propagators have the following spectral representations at fixed kaon
three-momentum $\vec{l}$
\begin{equation}
  \label{3.7}
  D_K(l_0,\, \vec{l}; \, \rho)= \int
  du^2 \frac{A_{K}(u,\, \vec{l}; \, \rho)}{l_0^2-\vec{l}^{\:2}-u^2+i\epsilon} 
\end{equation}
with
\begin{equation}
  \label{3.8}
  A_{K}(u,\, \vec{l};\, \rho )= \frac{ -\Im \,
  \Sigma_{K}(u,\, \vec{l};\, \rho)/\pi}{\left[u^2-m_K^2-\Re \,
  \Sigma_K(u,\,\vec{l};\,\rho)\right]^2+\left[ \Im
  \Sigma_K(u,\, \vec{l};\, \rho) \right]^2}.
\end{equation}
Once these spectral functions are determined at a given density for both
$K^+$, $K^0$ and $K^-$, $\overline{K}{}^0$, the $\phi$ meson self-energy is readily
calculated. For example, the $\phi \to K^+K^-$ in-medium self-energy for a
$\phi$ at rest becomes 
\begin{eqnarray}
\label{3.9}
&&\Pi_{\phi \to K^+K^-}(\omega,\, \vec{q}=0;\, \rho) =\nonumber \\ &&\frac{-i g^2}{6} \int
\frac{d^4l}{(2\pi)^4} \int du_+^2 \int du_-^2 \,\frac{ A_{K^+}(u_+,\,\vec{l};\, \rho)\,
A_{K^-}(u_-,\, \vec{l};\, \rho) \left[(2l_0-\omega)^2-4
\vec{l}^{\:2} \right]}{\left(l_0^2-\vec{l}^{\:2}-u_+^2+i \epsilon \right)\left((l_0-\omega)^2-\vec{l}^{\:2}-u_-^2
+i\epsilon\right)} \nonumber \\
&& \hspace*{4.3cm}+\, {\rm tadpole,}
\end{eqnarray}
where the tadpole part, not written explicitly, contributes only
to the real part of $\Pi_\phi$. 
Its imaginary part is now easily evaluated; for example:
\begin{equation}
\label{3.10}
\Im \, \Pi_{\phi \to K^+K^-}(\omega,\, \vec{q}=0;\, \rho) =\frac{- g^2}{96 \pi} \int du_+^2 \int
du_-^2 \, A_{K^+}(u_+,\, \vec{k}; \, \rho)\, A_{K^-}(u_-,\vec{k}; \,\rho ) \frac{\lambda^\frac{3}{2}(\omega^2,u_+^2,u_-^2)}{\omega^4}
\end{equation}
with $|\vec{k}|= \lambda^\frac{1}{2}(\omega^2,u_+^2,u_-^2)$, where
$\lambda(a,b,c) = a^2+b^2+c^2-2ab-2ac-2bc$ is the K\"allen function.

Let us now first discuss contributions to $\Im \, \Pi_\phi$ from s-wave $KN$
and ${\overline K}N$ interactions for which the leading process is illustrated
in Fig.1a. The corresponding in-medium kaon spectral
functions are determined using the coupled channels approach based on the
chiral SU(3) effective Lagrangian as described in refs.\cite{7,8}. This
approach successfully reproduces all available low-energy data of the $KN$ as
well as the coupled ${\overline K}N,\, \pi \Sigma,\, \pi \Lambda$ and $\eta
N,\, K \Lambda,\, K \Sigma$ systems. (For alternative approaches see refs. \cite{11,12,13}.)

For the $K^+$ and $K^0$ modes in matter one finds that the spectral functions
can be well approximated by a $\delta$-function with the free kaon mass
replaced by $m_{K^+}^\star(\rho)=m_{K^0}^\star(\rho)$: 
\begin{equation}
\label{3.13b}
A_{K^+,K^0}(u,\, \vec{l};\, \rho)=-\delta(u^2-m^{\star 2}_{K^+}(\rho)).
\end{equation}
At $\rho=\rho_0=0.17 \, \fm^{-3}$ we find  $m_{K^+}^\star\simeq 535$ MeV, nearly
independent of $\vec{l}$.

For the $K^-$ and $\overline{K}{}^0$ modes the spectrum generated by s-wave
${\overline K}N$ interactions has more interesting features. We give examples in
Fig.2 at $\rho=\rho_0$ for two energies $\omega$ of the external $\phi$ meson.
The ${\overline K}N$ system decays into $\pi \Lambda$ and $\pi \Sigma$. The
spectral functions $A_{K^-,\, \overline K{}^0}$ therefore have finite
widths. The second interesting point is the two-mode structure that appears at
finite $\overline K$
three-momentum $|\vec{k}|$. The low-energy ${\overline K} N$ s-wave interaction
is governed by the $\Lambda(1405)$ resonance. The response of nuclear matter to
$S=-1$ kaonic excitations therefore involves $\Lambda$(1405) particle - nucleon
hole states. Whereas the 
$\Lambda(1405)$ dissolves in matter for a $K^-$ at rest ($\vec{k}=0$) due to the
action of the Pauli principle \cite{8}, it reappears at finite $\vec{k}$. This is
seen in the two-peak structure of the spectral function, Fig.2, under
appropriate kinematical conditions. The lower mode can be identified with
a $\overline K$ (with its mass shifted downward) while the upper mode
corresponds to a $\Lambda (1405)$-hole excitation. When implemented in the $\phi$
meson self-energy, the upper ($\Lambda (1405)$) mode starts to become important
at energies $\omega \gsim 1.1$ GeV, above the free $\phi$ resonance. At these
energies the kaons move with relatively large momenta, and their scattering from
nucleons in nuclear matter is not much influenced by Pauli blocking effects.

Next we incorporate  p-wave $KN$ and ${\overline K} N$ interactions. They are
generated by the axial vector coupling terms in the chiral SU(3) effective
meson-baryon Lagrangian. The basic in-medium processes involve $\Lambda$- and
$\Sigma$-hole excitations, and we also include excitations of the spin-3/2
decuplet (the $\Sigma^*$). The octet and decuplet couplings are connected by
standard SU(3) relations. The p-wave $KN$ and ${\overline K}N$ contributions to
the $\phi N$ amplitude in leading order as sketched in Fig.1b and c. The
additional terms, Fig.1d and e, involve $\phi N\rightarrow KY$ contact
interactions representing short-range processes such as $K^*$ exchange. For baryon octet
($\Lambda$ and $\Sigma$) intermediate states one finds in the heavy-baryon
limit:
\begin{equation}
  \label{3.3}
  \Im T_{\phi N}^{\rm p-wave}(\omega,\,\vec{q}=0) = \frac{3}{4} (D-F)^2\, {\cal
  H}(\omega,M_\Sigma-M_N,m_K)+ \frac{1}{12} (D+3F)^2 \,{\cal
  H}(\omega,M_\Lambda-M_N,m_K),
\end{equation}
where $D=0.75$ and $F=0.51$ are the SU(3) axial coupling constants with
$g_A=D+F=1.26$. The function $\cal H$ (for $\omega>0$) is given by:
\begin{eqnarray}
  \label{3.4}
  {\cal H}(\omega,\Delta,m)& = & \nonumber \frac{g^2}{24 \pi
  f_\pi^2} \left[\frac{\sqrt{(\omega-\Delta)^2-m^2}}
  {(\omega^2-2\omega \Delta)^2} \left(3 \omega^4-4 \omega^2 m^2+4 m^4
  +\Delta (8 \omega m^2-12 \omega^3)\right. \right. \\
  \nonumber  && \hspace*{1.2cm}\left. \left.+\Delta^2 (16 \omega^2-8m^2 )-8 \Delta^3 \omega +4 \Delta^4
  \right)\right.\\  && \hspace*{1.2cm} \left.  -\frac{ \Delta (\omega^2-4
m^2)^\frac{3}{2}}{2 \omega \, (\omega^2-4 \Delta^2)^2} \left(3
\omega^2-8m^2-4 \Delta^2 \right) \right] .
\end{eqnarray}
We use $M_N=938\, MeV,\,M_\Lambda=1.116 \, \GeV$ and  $M_\Sigma=1.191 \,
\GeV$. The corresponding amplitude with a $\Sigma^*$ intermediate state has its
coupling scaled by the proper Clebsch-Gordan and SU(3) coefficients:
\begin{equation}
  \label{3.3}
  \Im T_{\phi N}^{p}(\omega) = \frac{3}{8} (D+F)^2\, {\cal
  H}(\omega,M_\Sigma^*-M_N,m_K).
\end{equation}
with $M_\Sigma^*=1.383 \, \GeV$.  In the actual calculations we have included
form factors at the axial kaon-baryon vertices. For simplicity we have used the
empirical nucleon axial form factor $G_A(t)=G_A(0)[1-q^2/\Lambda^2_A]^{-2}$
with $\Lambda_A=1.05$ GeV.
Note that the largest p-wave contribution comes from the intermediate
$\Lambda(1116)$. 

Summing all s- and p-wave contributions, we evaluate the imaginary
part of the effective, density dependent $\phi N$ amplitude for a $\phi$ meson
''at rest'':
\begin{equation}
  \label{3.13}
  \rho \, \Im T_{\phi N}^{K^+ K^-} (\omega,\, \vec{q}=0;\, \rho) = \Im
  \, \Pi_{\phi}^{\rm vac} (\omega,\, \vec{q}=0)-\Im \, \Pi_{\phi } (\omega,\,
  \vec{q}=0;\, \rho) .
\end{equation}
The real part of $T_{\phi N}$ is then determined by a subtracted dispersion
relation at $\vec{q}=0$:
\begin{equation}
  \label{3.14}
  \Re\, T_{\phi N} (\omega;\, \rho) = c_1+ \frac{\omega^2}{\pi}
  {\cal P} \int_{0}^\infty du^2 \frac{\Im T_{\phi N} (u;\,\rho)}{u^2
  (u^2-\omega^2)},
\end{equation}
where the subtraction constant $c_1=0$ is fixed by the Thomson limit of the
Compton amplitude involving the strange quark current (\ref{2.2}).

Our results for the real and imaginary parts of $T_{\phi N}$ at $\rho=\rho_0$
are shown in Fig.3. Note that the lower plateau in $\Im \, T_{\phi N}$  comes
mainly from the $\phi N \to K \Lambda$ channel, with an intermediate
${\overline K} N \to \Lambda$ p-wave coupling. The steep rise of $\Im \,
T_{\phi N}$ just above 0.9 GeV has two major contributions. About half of it
comes from $\phi N \to K {\overline K}N$ with s-wave interactions of the $K$ and
$\overline K$ in the nuclear medium. Roughly the other half originates from the
$\phi N \to K \Sigma^*$ process. The complexity of the low energy $\phi N$ dynamics
translates visibly into a highly structured pattern for  $\Re \, T_{\phi N}$
below and around the $\phi$ resonance. 

We mention that s-wave interactions of the ${\overline K}$ in the medium must
be treated to all orders in the density, even at small $\rho$. In contrast, the
iteration of p-wave interactions to higher orders in $\rho$ has a significant
effect only at energies above the $\phi$ resonance, at least for densities
$\rho \lsim \rho_0$. We estimate the uncertainties in the high energy part of
$\Im \, T_{\phi N} $ (at $\omega > m_\phi$) to be at the 20-30 \% level.
However, this uncertainty has very little influence on the in-medium spectral
function of the $\phi$ meson shown in Fig. 4a. In any case, the resulting
in-medium mass shift of the $\phi$ stays within 1 \% of its free mass.

The predicted spectrum of the strange current-current correlation function,
eq.(\ref{3.1}), is shown at $\rho=\rho_0$ in Fig.4 together with the real part
of the in-medium $\phi$ meson propagator,
\begin{equation}
  \label{3.16}
   D_\phi(\omega,\, \vec{q}=0;\, \rho) = \frac{1}{\omega^2-\stackrel{ \rm o
   }{m}_\phi^2-\Pi_\phi(\omega, \, \vec{q}=0;\, \rho)}.
\end{equation}
One observes a very small and insignificant shift of the resonance position (by about
1 \% of its free mass at $\rho=\rho_0$). The primary in-medium effect is the
broadening of the resonance due to inelastic $\phi N$ reactions. 
Its width at $\omega \simeq m_\phi$,
\begin{equation}
  \label{3.17}
  \Gamma_\phi=-\frac{\Im \, \Pi_\phi}{m_\phi},
\end{equation}
reaches $\Gamma_\phi \simeq 45 \, \MeV$ at $\rho=\rho_0$ and exceeds the free
(vacuum) width of 4 MeV by about an order of magnitude. The $\phi$ meson life time is
reduced to 
\begin{equation}
  \label{3.15}
  \tau_\phi=-\frac{1}{ \Gamma_\phi}\simeq 4.4\, {\rm fm/c \hspace*{0.5cm} at\; }\rho=\rho_0.
\end{equation}
Consequently, $\phi$ mesons implanted with small velocity in a
nucleus have a good chance to decay within nuclear dimensions.

{\bf Summary and conclusions}. We have evaluated the in-medium spectrum of the
$\phi$ meson taking the important s- and p-wave interactions of its $K
\overline K$ components with surrounding nuclear matter fully into account. At
normal nuclear matter density, $\rho=\rho_0$, we predict almost no mass
shift of the $\phi$ meson, while its width is expected to increase by about an
order of magnitude over its free width. The resulting short life time would
make it possible to observe in-medium modifications of a slowly moving $\phi$
within the diameter of a typical medium-heavy nucleus. Such considerations are
of interest for upcoming experiments using HADES at GSI. Finally we mention that
our calculated $\phi$ meson spectrum is consistent with the in-medium QCD sum rule
analysis of refs.\ \cite{2,6}.
\\
\\
\\
Acknowledgment:\\
We thank B. Friman and M. Lutz for fruitful discussions.

\newpage

\begin{center}
\unitlength=0.5mm
\begin{picture}(210,140)(20,20)
\SetWidth{2}   
\Photon(5,195)(30,170){3}{3}
\DashLine(30,170)(70,170){3}
\Photon(70,170)(95,195){3}{3}
\DashLine(30,170)(50,135){3}
\DashLine(50,135)(70,170){3}
\SetWidth{3}
\ArrowLine(25,110)(50,135)
\ArrowLine(50,135)(75,110)
\Text(36,69)[]{\large a}
\Text(36,128)[]{ K}
\Text(20,107)[]{ K}
\Text(18,138)[]{ $\phi$}
\Text(10,80)[]{ $N$}

\SetWidth{2}
\Photon(105,195)(130,170){3}{3}
\DashLine(130,170)(165,170){3}
\Photon(165,170)(190,195){3}{3}
\DashLine(130,170)(130,135){3}
\DashLine(165,170)(165,135){3}
\SetWidth{3}
\ArrowLine(105,110)(130,135)
\ArrowLine(165,135)(190,110)
\SetWidth{6}
\Line(130,135)(165,135)
\Text(103,69)[]{\large b}
\Text(103,128)[]{ K}
\Text(85,107)[]{ K}
\Text(120,107)[]{ K}
\Text(88,138)[]{ $\phi$}
\Text(68,80)[]{ $N$}
\Text(103,87)[]{$Y$}

\SetWidth{2}
\Photon(225,195)(250,170){3}{3}
\Photon(250,170)(275,195){3}{3}
\DashLine(250,170)(270,135){3}
\DashLine(230,135)(250,170){3}
\SetWidth{3}
\ArrowLine(205,110)(230,135)
\ArrowLine(270,135)(295,110)
\SetWidth{6}
\Line(230,135)(270,135)
\Text(176,69)[]{\large c}
\Text(161,108)[]{ K}
\Text(189,108)[]{ K}
\Text(170,138)[]{ $\phi$}
\Text(214,80)[]{ $N$}
\Text(176,87)[]{$Y$}

\SetWidth{2}
\Photon(35,60)(60,35){3}{3}
\Photon(100,35)(125,60){3}{3}
\DashCArc(80,35)(20,0,180){3}
\SetWidth{3}
\ArrowLine(35,10)(60,35)
\ArrowLine(100,35)(125,10)
\SetWidth{6}
\Line(60,35)(100,35)
\Text(55,0)[]{\large d}
\Text(55,47)[]{ K}
\Text(35,46)[]{ $\phi$}
\Text(21,15)[]{ $N$}
\Text(55,16)[]{$Y$}

\SetWidth{2}
\Photon(165,60)(190,35){3}{3}
\Photon(221,50)(245,74){3}{3}
\DashCArc(210,35)(20,0,180){3}
\SetWidth{3}
\ArrowLine(165,10)(190,35)
\ArrowLine(230,35)(255,10)
\SetWidth{6}
\Line(190,35)(230,35)
\Text(147,0)[]{\large e }
\Text(145,47)[]{ K}
\Text(126,46)[]{ $\phi$}
\Text(110,15)[]{ $N$}
\Text(147,16)[]{ $Y$}

\end{picture}\\
\end{center}
\vspace*{1.5cm}
Figure 1: Diagrams contributing to the leading s- and p-wave interactions of kaons in
nuclear matter. The hyperon intermediate states $Y$ include those of the baryon
octet and decuplet. \\

\begin{center}
\unitlength=1mm
\begin{picture}(100,80)
\put(0,0){\framebox(100,80){}}
\put(1,0){\makebox{\epsfig{file=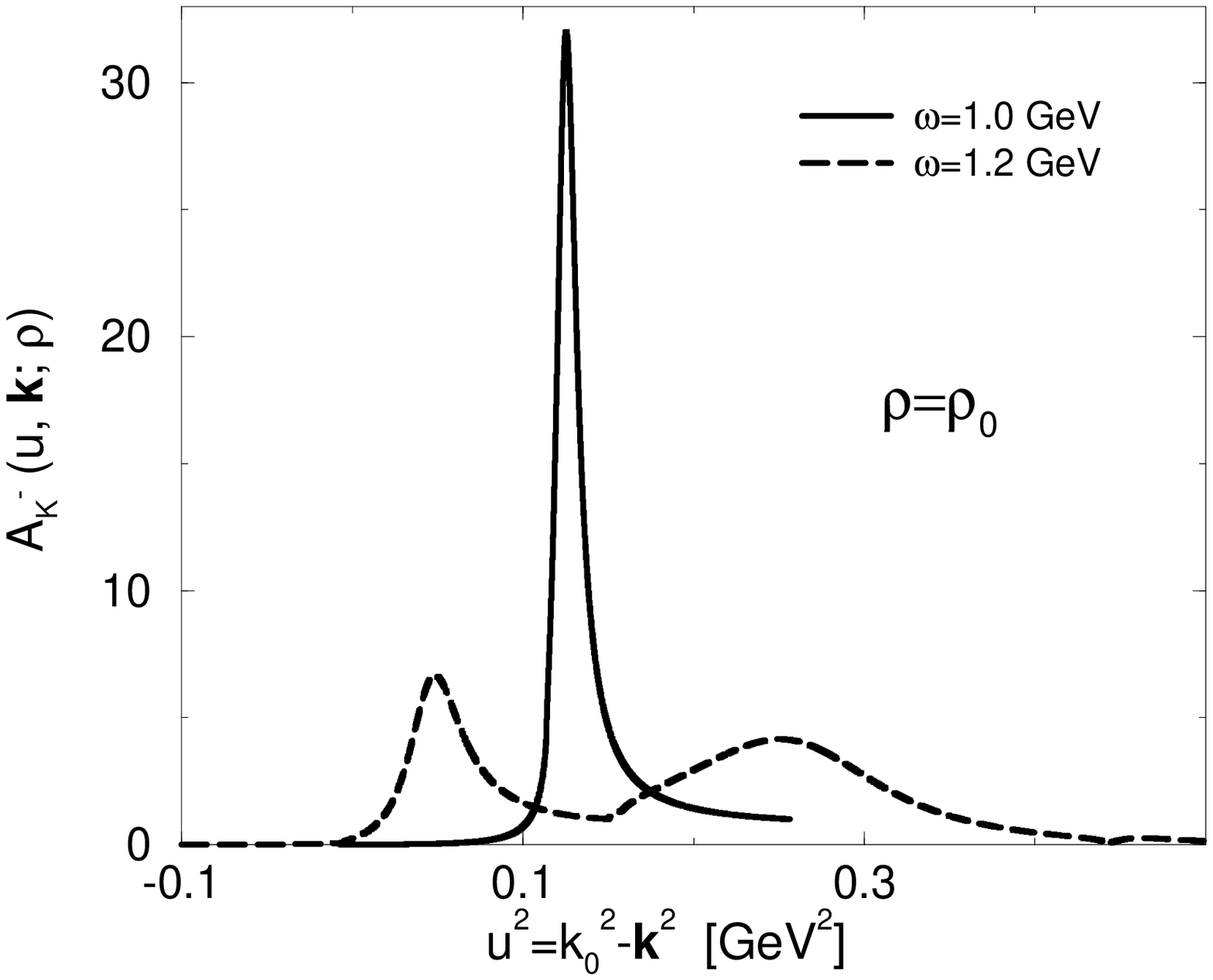,width=100mm}}}
\end{picture}
\end{center}
Figure 2: Spectral function $A_{K^-}$ of $K^-$ modes in nuclear matter at
density $\rho=\rho_0=0.17 \, \fm^{-3}$ for two energies ($\omega=1.0$ GeV and
$\omega=1.2$ GeV) of the primary $\phi$ meson (taken at rest), as a function of
the squared invariant mass $u^2=k_0^2-\vec{k}^2$. The calculation
includes s-wave $\overline K N$ interactions derived from chiral SU(3) dynamics
with coupled channels \cite{7,8}.

\begin{center}
\unitlength=1mm
\begin{picture}(100,80)
\put(0,0){\framebox(100,80){}}
\put(1,0){\makebox{\epsfig{file=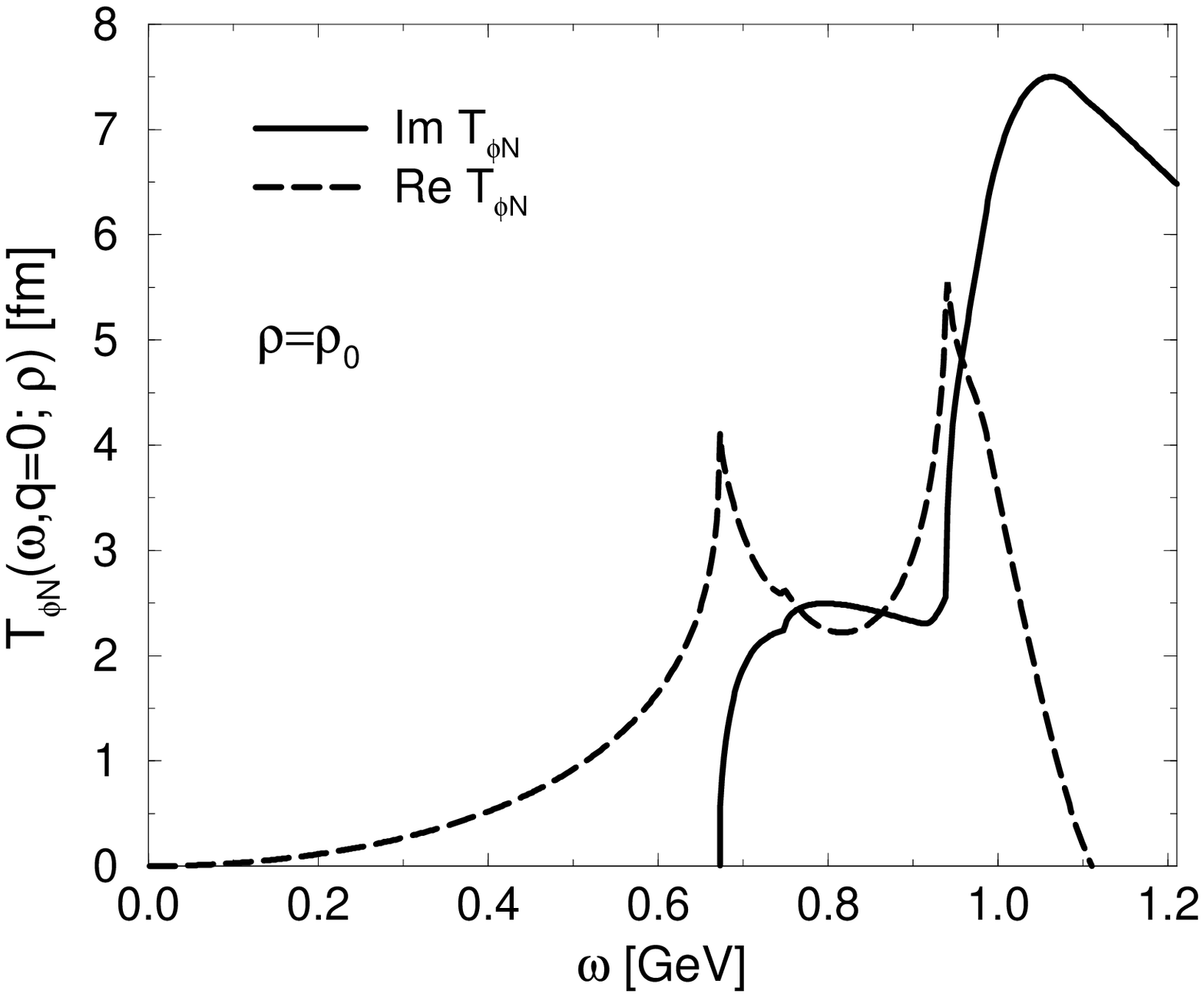,width=100mm}}}
\end{picture}
\end{center}
Figure 3: Real and imaginary parts of the effective $\phi N$ scattering
amplitude $T_{\phi N}(\omega, \vec{q}=0; \rho)$ in nuclear matter at density $\rho=\rho_0$. All s- and p-wave $KN$
and $\overline K N$ interactions are included.\\

\begin{center}
\unitlength=1mm
\begin{picture}(100,160)
\put(0,81){\line(1,0){100}}
\put(0,0){\framebox(100,160){}}
\put(1,0){\makebox{\epsfig{file=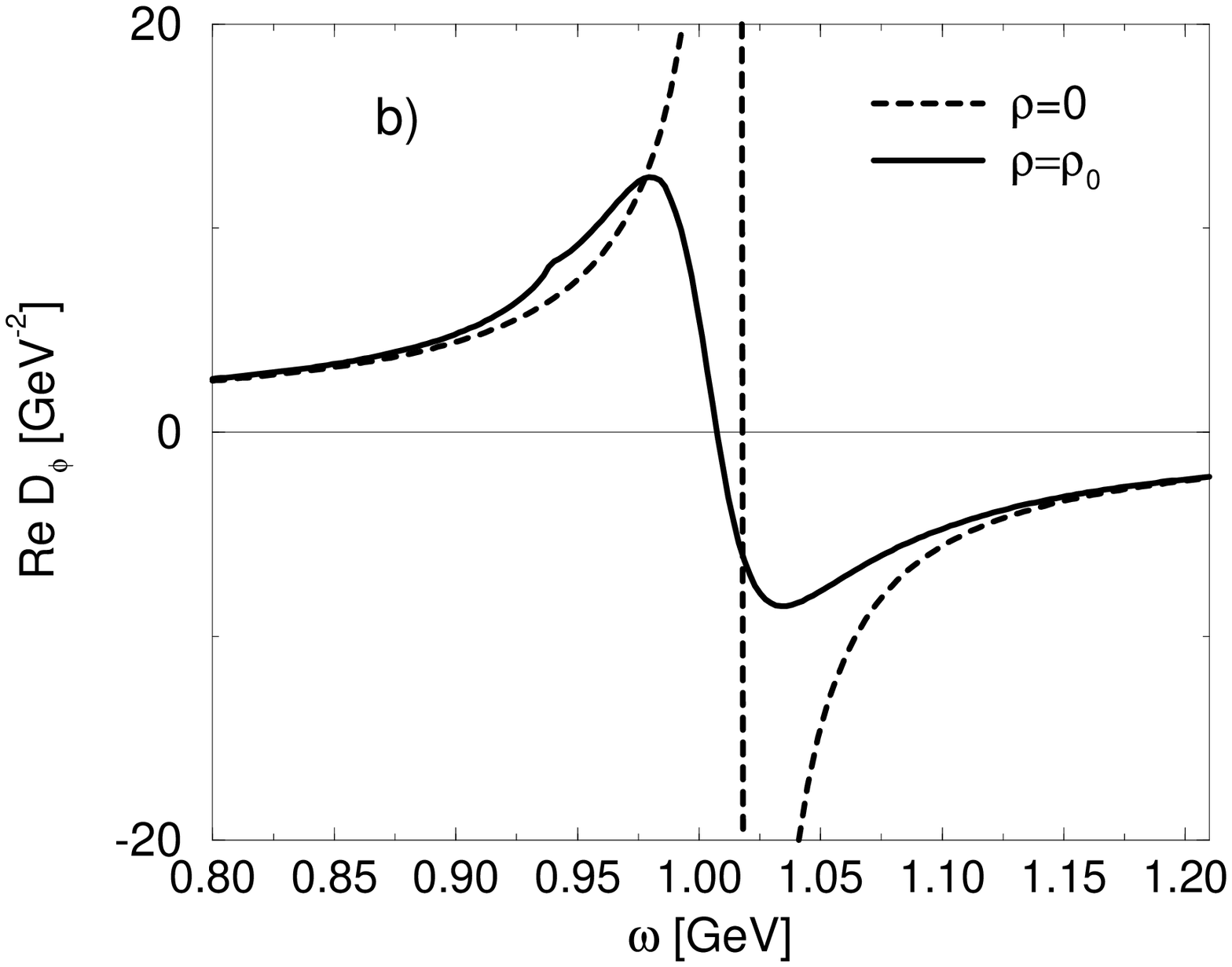,width=100mm}}}
\put(1,80){\makebox{\epsfig{file=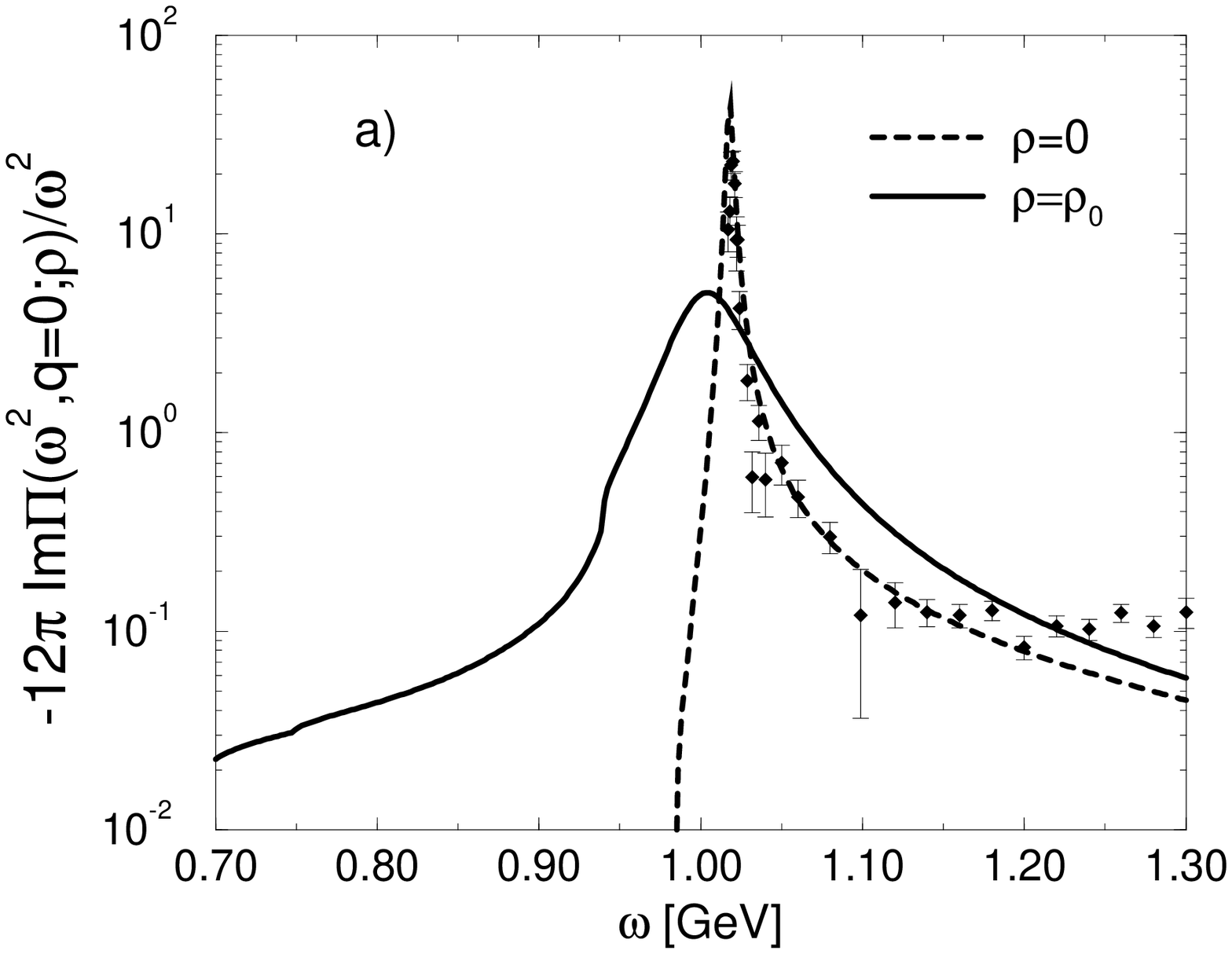,width=100mm}}}
\end{picture}
\end{center}
Figure 4: a) Spectrum of the strange quark current-current correlation
function in vacuum (dashed) and in nuclear matter at $\rho=\rho_0$ (solid
curve). The normalization is chosen such that the vacuum result can be compared
directly to the ratio $\sigma(e^+e^- \to K^+ K^-)/\sigma(e^+e^- \to \mu^+
\mu^-)$ for which data are taken from ref.\cite{10}.\\
b) Real part of the $\phi$ meson propagator in vacuum (dashed) and in nuclear
matter at $\rho=\rho_0$ (solid curve).\\


\begin{thebibliography}{99}

\bibitem{1} W. Sch\"on, H. Bokemeyer, W. Koenig and V. Metag, Acta
Phys. Polonica {\bf B 27} (1996) 2959.
\bibitem{9}E.V. Shuryak, Rev. Mod. Phys. {\bf{65}} (1993) 1
\bibitem{2}T. Hatsuda and S.H. Lee, Phys. Rev. {\bf{C 46}} (1992) R34;
M. Asakawa and C.M. Ko, Nucl. Phys. {\bf A 572} (1994) 732.
\bibitem{3} F. Klingl, N. Kaiser and W. Weise, Z. Phys. {\bf A 356} (1996) 193.
\bibitem{4} M. Herrmann, B.L. Friman and W. N\"orenberg, Nucl. Phys. {\bf A
560} (1993) 411.
\bibitem{5} F. Klingl and W. Weise, Nucl. Phys. {\bf A 606} (1996) 329.
\bibitem{6} F. Klingl, N. Kaiser and W. Weise, Nucl. Phys. {\bf A}(1997), in
print (hep-ph/9704398).
\bibitem{7} N. Kaiser, P.B Siegel and W. Weise, Nucl. Phys. {\bf A 594} (1995)
325; N. Kaiser, T. Waas and W.Weise, Nucl. Phys. {\bf A 612} (1997) 297.
\bibitem{8} T. Waas, N. Kaiser and W. Weise, Phys. Lett. {\bf B 379} (1996) 34;
T. Waas, M. Rho and W.Weise, Nucl. Phys. {\bf A 617} (1997) 449.
\bibitem{10} P.M. Ivanov et al., {Phys. Lett. }{\bf B 107} (1981) 297.
\bibitem{11} C.M. Ko, P. L\'evai, X. J. Qiu and C. T. Li, Phys. Rev. {\bf C 45} (1992) 1400.
\bibitem{12} C.S. Song, Phys. Lett. {\bf B 388} (1996) 141.
\bibitem{13} M. Asakawa and C. M. Ko, Phys. Lett. {\bf B 322} (1994) 33;
Phys. Rev.  {\bf C 50} (1994) 3064. 
\end{thebibliography}
\end{document}